\documentclass[english,aps,preprint,showpacs,preprintnumbers,amsmath,amssymb]{revtex4}
\usepackage[T1]{fontenc}
\usepackage[latin9]{inputenc}
\usepackage{color}
\usepackage{amsmath}
\usepackage{amssymb}
\makeatletter
\date{\today}
\makeatother
\usepackage{babel}
\begin{document}

\title{The Left-Right SU(3)$_{L}\otimes$SU(3)$_{R}\otimes$U(1)$_{X}$
Model \\
with Light, keV and Heavy Neutrinos}

\author{Alex G. Dias$^{1}$, C. A. de S. Pires$^{2}$, P. S. Rodrigues da
Silva$^{2}$}

\affiliation{\vspace{0.5cm}
 \\
 $^{1}$Centro de Ci\^encias Naturais e Humanas, Universidade Federal
do ABC,\\
 R. Santa Ad\'elia 166, Santo Andr\'e - SP, 09210-170, Brazil \vspace{0.3cm}
 \\
 $^{2}$ Departamento de F\'isica, Universidade Federal da Para\'iba,
Caixa Postal 5008, Jo\~ao Pessoa, PB 58051-970, Brazil.
 }

\date{\today}
\begin{abstract}
We construct a full left-right model for the electroweak interactions
based on the $SU(3)_{L}\otimes SU(3)_{R}\otimes U(1)_{X}$ gauge symmetry.
The fermion content of the model is such that anomaly cancellation
restricts the number of families to be a multiple of three. One of
the most important features of the model is the joint presence of
three light active neutrinos, three additional neutrinos at keV mass
scale, and six heavy ones with masses around\textbf{ $10^{11}$} GeV.
They form a well-motivated part of the spectrum in the sense that they
address challenging problems related to neutrino oscillation, warm
dark matter, and baryogenesis through leptogenesis.
\end{abstract}
\maketitle
The idea of explaining the observed maximal parity violation, incorporated
by the Standard Model, as a vacuum solution of a higher energy theory
manifestly invariant under left-right (LR) symmetry was initiated
by Pati and Salam~\cite{patisalam}. It was further developed, in
its minimal realization, with the $SU(2)_{L}\otimes SU(2)_{R}\otimes U(1)$
symmetry, in Ref.~\cite{senja-moha-75,mohapalbook}. As a byproduct
of such idea light left-handed neutrinos, as required by recent experiments
involving solar and atmospheric neutrino oscillations, arise through
the seesaw mechanism.

In this letter we construct and investigate a LR model based on $SU(3)_{L}\otimes SU(3)_{R}\otimes U(1)_{X}$
symmetry for the electroweak interactions, that we call 3L3R for short,
with a rich neutrino content that is well motivated by the problems
related to neutrino oscillation, warm dark matter, and baryogenesis
through leptogenesis. The matter representation content is defined
by requiring cancellation of all gauge anomalies involving three fermionic
families, concomitantly. In fact, the model can be extended to any
integer multiple of three fermionic families, although the asymptotic
freedom in the strong interaction sector restricts this number to
be exactly three. Therefore, the model we present here also naturally
embeds the $SU(3)_{L}\otimes U(1)_{N}$ models of \cite{svs} and
\cite{331nd}.

Let us start with the electric charge operator which involves a combination
of $SU(3)_{L,R}$ and $U(1)_{X}$ generators, 
\begin{equation}
Q=T_{L}^{3}+T_{R}^{3}+b(T_{L}^{8}+T_{R}^{8})+X\,.\label{eq:Q}
\end{equation}
\noindent 
Here the LR symmetry has been assumed so that the theory is invariant
by changing $L\rightleftarrows R$, and $b$ is a parameter defining
the charges of the fields forming the representations. The Standard
Model multiplets can be recovered by taking: $b=-1/\sqrt{3}$ or $b=\sqrt{3}$.
It happens that Eq.~(\ref{eq:Q}) leads to the following relation
involving the $SU(3)_{L,R}$ and $U(1)_{X}$ coupling constants, $g$
and $g_{X}$, respectively, and the electroweak mixing angle, 
\begin{equation}
\frac{g_{X}^{2}}{g^{2}}=\frac{sin^{2}\theta_{W}}{1-2\left(1+b^{2}\right)sin^{2}\theta_{W}}\,.
\label{eq:relcop}
\end{equation}
Remarkably, this relation implies that only $b=-1/\sqrt{3}$, which
restricts $sin^{2}\theta_{W}<3/8$, is viable in accordance with the
experimental fact that $sin^{2}\theta_{W_{(exp)}}\approx0.231$~\cite{pdg}.

Thus, the fermionic representation for this model is, omitting the
$SU(3)$ color quantum number,

\emph{for leptons:} \begin{eqnarray}
 &  & \Psi_{aL}\equiv\left[\nu_{La}\,\, e_{La}\,\, N_{La}\right]^{T}\sim\left(\mathbf{3,\,1,\,}-1/3\right)\,,\nonumber \\
 &  & \Psi_{aR}\equiv\left[\nu_{Ra}\,\, e_{Ra}\,\, N_{Ra}\right]^{T}\sim\left(\mathbf{1,\,3,\,}-1/3\right)\,,\label{eq:leptrip}\end{eqnarray}
 where $a=1,2,3$ is the family index, and

\emph{for quarks:} \begin{eqnarray}
 &  & Q_{mL}\equiv\left[d_{Lm}\,\, u_{Lm}\,\, d_{Lm}^{\prime}\right]^{T}\sim\left(\mathbf{3^{*},\,1,\,}0\right),\nonumber \\
 &  & Q_{3L}\equiv\left[u_{L3}\,\, d_{L3}\,\, u_{L3}^{\prime}\right]^{T}\sim\left(\mathbf{3,\,1,\,}1/3\right),\nonumber \\
 &  & Q_{mR}\equiv\left[d_{Rm\,\,}u_{Rm\,\,}d_{Rm}^{\prime}\right]^{T}\sim\left(\mathbf{1,\,3^{*},\,}0\right)\,,\nonumber \\
 &  & Q_{3R}\equiv\left[u_{R3}\,\, d_{R3}\,\, u_{R3}^{\prime}\right]^{T}\sim\left(\mathbf{1,\,3,\,}1/3\right)\,,\label{eq:qtrip}\end{eqnarray}
 with $m=1,2$. Observe that in this construction $Q_{mL}$ and $Q_{mR}$
are antitriplets of $SU(3)_{L}$ and $SU(3)_{R}$, respectively.
This implies that this model is free from the following anomalies:
\emph{\begin{eqnarray*}
I)\:\left[SU(3)_{_{color}}\right]^{2}U\left(1\right)_{X};\quad & II)\:[SU\left(3\right)_{L,R}]^{3};\\
III)\:[SU(3)_{L,R}]^{2}U\left(1\right)_{X};\quad\;\: & IV)\:[U\left(1\right)_{X}]^{3};\quad\;\;\\
V)\:[gravitational]^{2}U(1)_{X}\,.\end{eqnarray*}
 } The primed fields in Eq.~$\left(\ref{eq:qtrip}\right)$ are related
to new quarks with masses expected to be at the TeV scale as we shall
see.

We can now discuss the spontaneous breaking of the $SU(3)_{L}\times SU(3)_{R}\times U(1)_{X}$
to $U(1)_{em}$. We develop a minimal 3L3R model, meaning that we
introduce the minimum amount of scalar multiplets to give mass to
all fermionic fields through effective operators. Six scalar triplets
are taken into account, 
\begin{eqnarray}
 & \chi_{L}\equiv\left[\chi_{L}^{0}\,\,\chi_{L}^{-}\,\,\chi_{L}^{\prime0}\right]^{T}\sim\left(\mathbf{3,\,1,\,}-1/3\right), & \chi_{R}\equiv\left[\chi_{R}^{0}\,\,\chi_{R}^{-}\,\,\chi_{R}^{\prime0}\right]^{T}\sim\left(\mathbf{1,\,3,\,}-1/3\right)\nonumber \\
 & \eta_{L}\equiv\left[\eta_{L}^{0}\,\,\eta_{L}^{-}\,\,\eta_{L}^{\prime0}\right]^{T}\sim\left(\mathbf{3,\,1,\,}-1/3\right),\; & \eta_{R}\equiv\left[\eta_{R}^{0}\,\,\eta_{R}^{-}\,\,\eta_{R}^{\prime0}\right]^{T}\sim\left(\mathbf{1,\,3,\,}-1/3\right)\nonumber \\
 & \rho_{L}\equiv\left[\rho_{L}^{+}\,\,\rho_{L}^{0}\,\,\rho_{L}^{\prime+}\right]^{T}\sim\left(\mathbf{3,\,1,\,}2/3\right),\;\;\;\; & \rho_{R}\equiv\left[\rho_{R}^{+}\,\,\rho_{R}^{0}\,\,\rho_{R}^{\prime+}\right]^{T}\sim\left(\mathbf{1,\,3,\,}2/3\right).
 \label{eq:esctrip}
 \end{eqnarray}
 It is assumed here that only six of the ten neutral scalars develop
a nonzero vacuum expectation value (VEV): 
\begin{eqnarray}
 & \langle\chi_{L}\rangle=\frac{1}{\sqrt{2}}\left[0\,,\,0\,,\, v_{\chi_{L}^{\prime}}\right]^{T}, & \langle\chi_{R}\rangle=\frac{1}{\sqrt{2}}\left[0\,,\,0\,,\, v_{\chi_{R}^{\prime}}\right]^{T}\nonumber \\
 & \langle\eta_{L}\rangle=\frac{1}{\sqrt{2}}\left[v_{\eta_{L}}\,,\,0\,,\,0\right]^{T}, & \langle\eta_{R}\rangle=\frac{1}{\sqrt{2}}\left[v_{\eta_{R}}\,,\,0\,,\,0\right]^{T}\nonumber \\
 & \langle\rho_{L}\rangle=\frac{1}{\sqrt{2}}\left[0\,,\, v_{\rho_{L}}\,,\,0\right]^{T}, & \langle\rho_{R}\rangle=\frac{1}{\sqrt{2}}\left[0\,,\, v_{\rho_{R}}\,,\,0\right]^{T},\label{eq:vevs}
 \end{eqnarray}
which must be consistent with the minimum of the scalar potential.
To avoid unnecessary complications in the construction of the model
and for future convenience, when analyzing the neutrinos aspects,
we impose the discrete symmetry \textbf{\begin{eqnarray}
\rho_{L,R} & \rightarrow & -\rho_{L,R}\nonumber \\
\chi_{L,R} & \rightarrow & -\chi_{L,R},\label{eq:disym}\end{eqnarray}
}with all remaining fields transforming trivially. The general scalar
potential invariant under LR, gauge, and the discrete symmetry, can
be written as \textbf{\begin{align}
V & =\sum_{I}\mu_{I}^{2}\left(\Phi_{IL}^{\dagger}\Phi_{IL}+\Phi_{IR}^{\dagger}\Phi_{IR}\right)+f\varepsilon^{ijk}\left(\chi_{Li}\eta_{Lj}\rho_{Lk}+\chi_{Ri}\eta_{Rj}\rho_{Rk}+H.c\right)\nonumber \\
 & +\sum_{I}\lambda_{I}\left[(\Phi_{IL}^{\dagger}\Phi_{IL})^{2}+(\Phi_{IR}^{\dagger}\Phi_{IR})^{2}\right]+\sum_{I,J}\alpha_{IJ}(\Phi_{IL}^{\dagger}\Phi_{IL})(\Phi_{JR}^{\dagger}\Phi_{JR})\nonumber \\
 & +\frac{1}{2}\sum_{J\neq I}\lambda_{IJ}\left[(\Phi_{IL}^{\dagger}\Phi_{IL})(\Phi_{JL}^{\dagger}\Phi_{JL})+(\Phi_{IR}^{\dagger}\Phi_{IR})(\Phi_{JR}^{\dagger}\Phi_{JR})\right]\nonumber \\
 & +\frac{1}{2}\sum_{J\neq I}\kappa_{IJ}\left[(\Phi_{IL}^{\dagger}\Phi_{JL})(\Phi_{JL}^{\dagger}\Phi_{IL})+(\Phi_{IR}^{\dagger}\Phi_{JR})(\Phi_{JR}^{\dagger}\Phi_{IR})\right]\nonumber \\
 & +\frac{1}{2}\sum_{J\neq I}\kappa_{IJ}^{\prime}\left[(\Phi_{IL}^{\dagger}\Phi_{JL})(\Phi_{JR}^{\dagger}\Phi_{IR})+(\Phi_{IR}^{\dagger}\Phi_{JR})(\Phi_{JL}^{\dagger}\Phi_{IL})\right]\label{eq:pot}\end{align}
}with the three components object $\Phi\equiv\left(\chi,\:\eta,\:\rho\right)$.\textbf{
}The minimum condition for this potential leads to the constraint
equations, {\small \begin{eqnarray}
\lambda_{3}v_{\rho_{L}}^{2}+\frac{1}{2}\left(\lambda_{13}v_{\chi_{L}^{\prime}}^{2}+\lambda_{23}v_{\eta_{L}}^{2}+\alpha_{33}v_{\rho_{R}}^{2}+\alpha_{13}v_{\chi_{R}^{\prime}}^{2}+\alpha_{23}v_{\eta_{R}}^{2}+\sqrt{2}f\frac{v_{\eta_{L}}v_{\chi_{L}^{\prime}}}{v_{\rho_{L}}}\right) & = & -\mu_{\rho}^{2},\nonumber \\
\lambda_{3}v_{\rho_{R}}^{2}+\frac{1}{2}\left(\lambda_{13}v_{\chi_{R}^{\prime}}^{2}+\lambda_{23}v_{\eta_{R}}^{2}+\alpha_{33}v_{\rho_{L}}^{2}+\alpha_{13}v_{\chi_{L}^{\prime}}^{2}+\alpha_{23}v_{\eta_{L}}^{2}+\sqrt{2}f\frac{v_{\eta_{R}}v_{\chi_{R}^{\prime}}}{v_{\rho_{R}}}\right) & = & -\mu_{\rho}^{2},\nonumber \\
\lambda_{1}v_{\chi_{L}^{\prime}}^{2}+\frac{1}{2}\left(\lambda_{12}v_{\eta_{L}}^{2}+\lambda_{13}v_{\rho_{L}}^{2}+\alpha_{11}v_{\chi_{R}^{\prime}}^{2}+\alpha_{12}v_{\eta_{R}}^{2}+\alpha_{13}v_{\rho_{R}}^{2}+\sqrt{2}f\frac{v_{\eta_{L}}v_{\rho_{L}}}{v_{\chi_{L}^{\prime}}}\right) & = & -\mu_{\chi}^{2},\nonumber \\
\lambda_{1}v_{\chi_{R}^{\prime}}^{2}+\frac{1}{2}\left(\lambda_{12}v_{\eta_{R}}^{2}+\lambda_{13}v_{\rho_{R}}^{2}+\alpha_{11}v_{\chi_{L}^{\prime}}^{2}+\alpha_{12}v_{\eta_{L}}^{2}+\alpha_{13}v_{\rho_{L}}^{2}+\sqrt{2}f\frac{v_{\eta_{R}}v_{\rho_{R}}}{v_{\chi_{R}^{\prime}}}\right) & = & -\mu_{\chi}^{2},\nonumber \\
\lambda_{2}v_{\eta_{L}}^{2}+\frac{1}{2}\left(\lambda_{12}v_{\chi_{L}^{\prime}}^{2}+\lambda_{23}v_{\rho_{L}}^{2}+\alpha_{22}v_{\eta_{R}}^{2}+\alpha_{12}v_{\chi_{R}^{\prime}}^{2}+\alpha_{23}v_{\rho_{R}}^{2}+\sqrt{2}f\frac{v_{\chi_{L}^{\prime}}v_{\rho_{L}}}{v_{\eta_{L}}}\right) & = & -\mu_{\eta}^{2},\nonumber \\
\lambda_{2}v_{\eta_{R}}^{2}+\frac{1}{2}\left(\lambda_{12}v_{\chi_{R}^{\prime}}^{2}+\lambda_{23}v_{\rho_{R}}^{2}+\alpha_{22}v_{\eta_{L}}^{2}+\alpha_{12}v_{\chi_{L}^{\prime}}^{2}+\alpha_{23}v_{\rho_{L}}^{2}+\sqrt{2}f\frac{v_{\chi_{R}^{\prime}}v_{\rho_{R}}}{v_{\eta_{R}}}\right) & = & -\mu_{\eta}^{2}\label{eq:vinc}\end{eqnarray}
} With this, it is easy to check that no constraint emerges to force
some of the VEVs to be identically zero, which suits our minimal implementation
of spontaneous symmetry breaking and mass generation.

The 3L3R will be spontaneously broken to $U(1)_{em}$ as follows:
the VEVs, \textcolor{black}{$v_{\chi_{R}^{\prime}}$, $v_{\rho_{R}}$
and $v_{\eta_{R}}$ first realize }$SU(3)_{L}\otimes SU(3)_{R}\otimes U(1)_{X}/SU(3)_{L}\otimes U(1)_{N}$;
while reduction to the electromagnetic factor $SU(3)_{L}\otimes U(1)_{N}/U(1)_{em}$
is then reached through the VEV's $v_{\chi_{L}^{\prime}}$, $v_{\rho_{L}}$
and $v_{\eta_{L}}$. This last symmetry reduction should be realized
with the VEV $v_{\chi_{L}^{\prime}}$ leading to $SU(3)_{L}\otimes U(1)_{N}/SU(2)_{L}\otimes U(1)_{Y}$
and, then, the VEV's $v_{\rho_{L}}$ and $v_{\eta_{L}}$ leading to
$SU(2)_{L}\otimes U(1)_{Y}/U(1)_{em}$. Therefore, it is reasonable
to have $v_{\chi_{L}^{\prime}}$ $>$$v_{\rho_{L}}$,$v_{\eta_{L}}$.
It is specially interesting that keV mass neutrinos in the 3L3R model
suggest $v_{\chi_{L}^{\prime}}$ to be at the TeV scale, making it possible
for the new particles of the model be produced in the present particle
colliders. As we will see later, breaking of the 3L3R symmetry at
a very high energy scale to $SU(3)_{L}\otimes U(1)_{N}$ yields a
successful seesaw mechanism for the neutrinos.

We can now make certain that the vacuum alignment taken above is indeed
a physical solution, by investigating the scalar mass spectrum of
the model arising from Eq.~(\ref{eq:pot}), also taking into account
Eq.~(\ref{eq:vinc}). Upon scalars mass matrix diagonalization, we
see that from the 36 degrees of freedom in Eq.~(\ref{eq:esctrip}),
16 are Goldstone bosons which form part of the degrees of freedom
of the massive gauge bosons. In the scalar particles spectrum, there
are six CP even and two CP odd scalar fields, two non-Hermitian neutral
scalar fields, and four charged scalar fields. Half of these fields
are directly related to the energy scales \textcolor{black}{$v_{\rho_{R}}$,
$v_{\chi_{R}^{\prime}}$, and $v_{\eta_{R}}$ that we designate as
}R-energy scales of the model\textcolor{black}{{} (we also designate
}$v_{\chi_{L}^{\prime}}$, $v_{\rho_{L}}$, and $v_{\eta_{L}}$\textcolor{black}{{}
as the L-energy scales). It is assumed here that these R-energy scales
are very high, so that only the particles directly related to the
left scales belong to the low energy spectrum.} It can be argued that
there are regions in the parameter space of the potential in Eq.~(\ref{eq:pot})
which lead to a physical spectrum for the scalar fields. For example,
we can assume that the couplings among L and R scalar triplets are
very small, i. e., $\alpha_{IJ},\:$$\kappa_{IJ}^{\prime}\thickapprox0$.
In this case the scalar fields related to the L-energy scales are
the same as those in Ref. \cite{331nd}. The squared masses of the
scalar fields which remain in the spectrum are all positive as it
is required to have a stable minimum point in the potential with
the VEV configuration we have assumed. This was studied in Ref. \cite{ppdm331}
for a model which has the same scalar sector, like the low energy
effective theory derived from the 3L3R model. The scalar particles, as
well as the new gauge bosons and fermions related to the \textcolor{black}{L-energy
scales} may potentially be probed in the LHC.

Concerning the gauge bosons sector, the spectrum involves the photon
and the other 16 massive gauge bosons where eight of them are charged
($W_{L}^{\pm}\,,\, V_{L}^{\pm}\,,\, W_{R}^{\pm}\,,\, V_{R}^{\pm}$),
four are neutral but non-Hermitian ($U_{L}^{0}\,,\, U_{L}^{0\dagger}\,,\, U_{R}^{0}\,,\, U_{R}^{0\dagger}$),
and the other four are neutral ($Z_{L}^{0}\,,\, Z_{L}^{0\prime}\,,\, Z_{R}^{0}\,,\, Z_{R}^{0\prime}$).
Taking into account that $SU(3)_{L}\otimes SU(3)_{R}\otimes U(1)_{X}/SU(3)_{L}\otimes U(1)_{N}$
is linked to a very high energy scale the gauge bosons associated
with the right-handed part of the model decouple from the spectrum associated
with the left-handed part which, on the other hand, is basically the
set of gauge bosons characteristic of the $SU(3)_{L}\otimes U(1)_{N}$
model \cite{331nd,economical331model}. The gauge bosons $Z_{L}^{0}$
and $W_{L}^{\pm}$ recover the properties of the standard gauge bosons
$Z^{0}$ and $W^{\pm}$, and $V_{L}^{\pm}$, $U_{L}^{0}$, and $U_{L}^{0\dagger}$
are all mainly linked to the $v_{\chi_{L}^{\prime}}$ scale. For more
specific details of these left gauge bosons we refer the readers to 
Ref.~\cite{long}.

It should be observed that it is still possible to construct an analogous
LR model like we propose here, but considering only four scalar triplets 
(two left scalar triplets and two right scalar triplets). Although
the fact that such construction leads to a simpler scalar sector it
seems it is not possible to generate all fermion masses exclusively
by means of VEVs. In this case additional mechanisms have to be implemented,
as, for example, the radiative mass corrections \cite{economical331model,donglong2008}.

Now let us move to the generation of the fermion masses. Such masses
can emerge uniquely through effective operators, once we have only
scalar triplets in the present model. Considering dimension-5 operators
we have for the leptons \begin{eqnarray}
\mathcal{L}_{eff}^{l} & = & \frac{h_{ab}^{l}}{\Lambda_{D}}(\overline{\Psi}_{aL}\rho_{L})(\rho_{R}^{\dagger}\Psi_{bR})+\frac{g_{ab}^{D}}{\Lambda_{D}}(\overline{\Psi}_{aL}\chi_{L})(\chi_{R}^{\dagger}\Psi_{bR})+\frac{y_{ab}^{D}}{\Lambda_{D}}(\overline{\Psi}_{aL}\eta_{L})(\eta_{R}^{\dagger}\Psi_{bR})\nonumber \\
 & + & \frac{g_{ab}^{M}}{\Lambda_{M}}\left[\left(\overline{\left(\Psi_{aL}\right)^{c}}\chi_{L}^{*}\right)\left(\chi_{L}^{\dagger}\Psi_{bL}\right)+\left(\overline{\left(\Psi_{aR}\right)^{c}}\chi_{R}^{*}\right)\left(\chi_{R}^{\dagger}\Psi_{bR}\right)\right]\nonumber \\
 & + & \frac{y_{ab}^{M}}{\Lambda_{M}}\left[\left(\overline{\left(\Psi_{aL}\right)^{c}}\eta_{L}^{*}\right)\left(\eta_{L}^{\dagger}\Psi_{bL}\right)+\left(\overline{\left(\Psi_{aR}\right)^{c}}\eta_{R}^{*}\right)\left(\eta_{R}^{\dagger}\Psi_{bR}\right)\right]+H.c,\label{eq:efoplep}\end{eqnarray}
 and for the quarks\begin{align}
\mathcal{L}_{eff}^{q} & =\frac{h_{mn}^{u}}{\Lambda_{D}}(\overline{Q}_{mL}\rho_{L}^{*})(\rho_{R}^{T}Q_{nR})+\frac{h_{33}^{\chi u}}{\Lambda_{D}}(\overline{Q}_{3L}\chi_{L})(\chi_{R}^{\dagger}Q_{3R})\nonumber \\
 & +\frac{h_{33}^{\eta u}}{\Lambda_{D}}(\overline{Q}_{3L}\eta_{L})(\eta_{R}^{\dagger}Q_{3R})+\frac{h_{mn}^{\chi d}}{\Lambda_{D}}(\overline{Q}_{mL}\chi_{L}^{*})(\chi_{R}^{T}Q_{nR})\nonumber \\
 & +\frac{h_{mn}^{\eta d}}{\Lambda_{D}}(\overline{Q}_{mL}\eta_{L}^{*})(\eta_{R}^{T}Q_{nR})+\frac{h_{33}^{d}}{\Lambda_{D}}(\overline{Q}_{3L}\rho_{L})(\rho_{R}^{\dagger}Q_{3R})\nonumber \\
 & +\frac{h_{m3}^{u}}{\Lambda_{D}}\left[(\overline{Q}_{mL}\rho_{L}^{*})(\chi_{R}^{\dagger}Q_{3R})+(\overline{Q}_{mR}\rho_{R}^{*})(\chi_{L}^{\dagger}Q_{3L})\right]\nonumber \\
 & +\frac{h_{m3}^{\chi d}}{\Lambda_{D}}\left[(\overline{Q}_{mL}\chi_{L}^{*})(\rho_{R}^{\dagger}Q_{3R})+(\overline{Q}_{mR}\chi_{R}^{*})(\rho_{L}^{\dagger}Q_{3L})\right]+H.c\label{eq:efopq}\end{align}
These operators have introduced two new scales into the model, $\Lambda_{D}$,
which is associated with Dirac mass terms, and also, $\Lambda_{M}$,
which is associated with Majorana mass terms. Thus, having in mind
a kind of seesaw mechanism behind fermion masses, it is plausible
to expect that $\Lambda_{M}>>\Lambda_{D}$, and we choose $\Lambda_{M}$
to be the Planck scale and $\Lambda_{D}$ some unification scale.
Moreover, since all Dirac masses of quarks and leptons involve a ratio
between the R-energy scales and $\Lambda_{D}$, we can infer that
it is not inappropriate to assume $v_{\rho_{R}}\approx v_{\eta_{R}}\approx v_{\chi_{R}^{\prime}}\approx\Lambda_{D}$.
A similar approach for giving mass to the fermions like the way we
use here was shown in a $SU(2)_{L}\otimes SU(2)_{R}$ in Ref.~\cite{maLR}.

In what concerns the charged leptons their masses are obtained from
the first operator in Eq.~(\ref{eq:efoplep}), giving $m_{l}\approx\frac{h^{l}}{2}v_{\rho_{L}}$.
 The known quarks have their masses linked to the $v_{\rho_{L}}$,
$v_{\eta_{L}}$ scales, while the new exotic fermions have their masses
linked to the $v_{\chi_{L}^{\prime}}$ scale, mainly. Besides, there
is not any charged fermion with mass at the scale $\Lambda_{D}$.\textbf{
}As is reasonable to suppose, with $v_{\chi_{L}^{\prime}}$ situated
at TeV scale, the new quarks may be probed at the LHC.\textbf{ }Moreover,
there is also a mixing among new quarks and the standard ones, making
the quark sector of the 3L3R model more complex than that in the Standard
Model. The quark sector of the 3L3R model will be developed elsewhere.

So far we have implemented a genuine LR model with a minimal content
and also successfully recovered the important features of the Standard
Model. Besides, there are new extra features which are interesting
byproducts of a $SU(3)_{L}\otimes U(1)_{N}$ electroweak gauge model.

Next we deal with the neutrino physics showing another peculiar characteristic
that, as far as we know, is only present in this 3L3R model. The dimension-5
effective operators in Eq. $\left(\ref{eq:efoplep}\right)$ provide
Majorana as well as Dirac mass terms for the neutrinos. Defining the
basis $(\Psi_{L}\,,\,(\Psi_{R})^{C})$, where $\Psi_{L}=(\nu_{L}\,,\, N_{L})$
and $(\Psi_{R})^{C}=((\nu_{R})^{C}\,,\,(N_{R})^{C})$, we obtain \begin{eqnarray}
M_{\nu}=\left(\begin{array}{cc}
M_{L} & M_{D}\\
M_{D}^{T} & M_{R}\end{array}\right),\label{generalmatrix}\end{eqnarray}
which is a $12\times12$ matrix constituted by four blocks of $6\times6$
matrices, where $M_{D}$ is generated by the second and the third
terms in Eq.~(\ref{eq:efoplep}), \begin{eqnarray}
M_{D}=\frac{1}{2\Lambda_{D}}\left(\begin{array}{cc}
y^{D}v_{\eta_{L}}v_{\eta_{R}} & 0\\
0 & g^{D}v_{\chi_{L}^{\prime}}v_{\chi_{R}^{\prime}}\end{array}\right),\label{MD}\end{eqnarray}
 the fourth and the fifth terms generates $M_{L}$, \begin{eqnarray}
M_{L}=\frac{1}{\Lambda_{M}}\left(\begin{array}{cc}
y^{M}v_{\eta_{L}}^{2} & 0\\
0 & g^{M}v_{\chi_{L}^{\prime}}^{2}\end{array}\right),\label{ML}\end{eqnarray}
 and $M_{R}$, \begin{eqnarray}
M_{R}=\frac{1}{\Lambda_{M}}\left(\begin{array}{cc}
y^{M}v_{\eta_{R}}^{2} & 0\\
0 & g^{M}v_{\chi_{R}^{\prime}}^{2}\end{array}\right).\label{MR}\end{eqnarray}

 Here, $g^{D}$, $g^{M}$, $y^{D}$ and $y^{M}$ are $3\times3$ matrices
whose elements are taken real for simplicity. The assumption that
R-energy scales are much higher than the L-energy scales leads to
a hierarchy among the magnitudes of the elements in the matrices above.
$M_{R}$ involves the ratios $v_{\eta_{R}}^{2}/\Lambda_{M}$, $v_{\chi_{R}^{\prime}}^{2}/\Lambda_{M}$,
while $M_{D}$ involves ratios $v_{\eta_{L}}v_{\eta_{R}}/\Lambda_{D}$,
$v_{\chi_{L}^{\prime}}v_{\chi_{R}^{\prime}}/\Lambda_{D}$ and $M_{L}$
involves ratios $v_{\eta_{L}}^{2}/\Lambda_{M}$, $v_{\chi_{L}^{\prime}}^{2}/\Lambda_{M}$.
As $\Lambda_{M}>>\Lambda_{D}$, in comparing the order of magnitude
of these block matrices, we can safely neglect $M_{L}$ in Eq.~(\ref{generalmatrix}).
In view of this, after diagonalization of $M_{\nu}$, we obtain a
{\bf $6\times6$ } block diagonal matrix with the following mass matrix elements,
\begin{eqnarray}
M_{\nu_{L}^{\prime}}\approx-M_{D}(M_{R})^{-1}M_{D}^{T}\,\,\,\,\,\,\,\,\,\,\,\, M_{\nu_{R}}=M_{R}.\label{numatrix1}\end{eqnarray}
The resulting mixing among the components of $(\Psi_{R})^{C}$ and
$\Psi_{L}$ is of the order of the matrix elements in $M_{D}M_{R}^{-1}$,
which are proportional to the ratio $\Lambda_{M}/\Lambda_{D}^{2}$
times $v_{\eta_{L}}$ or $v_{\chi_{L}^{\prime}}$. Thus, $\Psi_{L}$
practically decouples from $(\Psi_{R})^{C}$.

The explicit form for $M_{\nu_{L}^{\prime}}$ is approximately \begin{eqnarray}
 &  & M_{\nu_{L}^{\prime}}\approx-\frac{\Lambda_{M}}{4\Lambda_{D}^{2}}\left(\begin{array}{cc}
y^{D}(y^{M})^{-1}(y^{D})^{T}v_{\eta_{L}}^{2} & 0\\
0 & g^{D}(g^{M})^{-1}(g^{D})^{T}v_{\chi_{L}^{\prime}}^{2}\end{array}\right)\label{intermediatematrix}\end{eqnarray}
 A nice result of the alignment of the VEV's is that the matrix
$M_{\nu_{L}^{\prime}}$ gets block diagonal, with the first block
being the Majorana mass matrix of the standard left-handed neutrinos $\nu_L$, and the second one being the Majorana mass matrix of the new left-handed neutrino $N_L$,
\begin{eqnarray}
 &  & M_{\nu_{L}}\approx-\frac{\Lambda_{M}v_{\eta_{L}}^{2}}{4\Lambda_{D}^{2}}y^{D}(y^{M})^{-1}(y^{D})^{T},\nonumber \\
 &  & M_{N_{L}}\approx-\frac{\Lambda_{M}v_{\chi_{L}^{\prime}}^{2}}{4\Lambda_{D}^{2}}g^{D}(g^{M})^{-1}(g^{D})^{T}.\label{3x3massmatrix}\end{eqnarray}
 On taking the values $\Lambda_{M}=10^{19}$~GeV, $\Lambda_{D}=10^{15}$~GeV,
$v_{\eta_{L}}=20$~GeV, and $v_{\chi_{L}^{\prime}}=2\times10^{3}$~GeV,
the model predicts the following order of magnitude for the masses
of the left-handed neutrinos: 
\begin{eqnarray}
 &  & M_{\nu_{L}}\approx y^{D}(y^{M})^{-1}(y^{D})^{T}\mbox{eV},\nonumber \\
 &  & M_{N_{L}}\approx10g^{D}(g^{M})^{-1}(g^{D})^{T}\mbox{keV},
 \label{masspredictions}
 \end{eqnarray}
 We call the attention to two nice results here. First, the order
of magnitude of the mass matrices above where we have eV for the standard
left-handed neutrinos, which is in agreement with solar and atmospheric
neutrino oscillation experiments. The typical order of magnitude for
the neutrino masses that convey an explanation for these experiments
seems to be in the range $10^{-3}-10^{-2}$ eV\textbf{ }\cite{moha}.
Also, we have a keV mass scale for the $N_{L}$ neutrinos, which falls
in the range of values required by warm dark matter candidates \cite{kusenko2009}.
The second nice result is that $N_{L}$ does not mix with $\nu_{L}$.
In view of this, we have that the lightest $N_{L}$ turns out to be\textbf{
}a stable particle. Moreover, one problem that could jeopardize sterile
neutrino as a warm dark matter candidate is its decay in x-ray\textbf{
$\nu_{S}\rightarrow\gamma+\nu$}. The advantage of $N_{L}$ as dark
matter candidate is that, due to the absence of mixing among $N_{L}$
and $\nu_{L}$, it gets free from the x-ray constraint, i.e. we do
not have the decay $N_{L}\rightarrow\gamma+\nu_{L}$. All this makes
$N_{L}$ a safe warm dark matter candidate.

One might ask about the hierarchy that emerges concerning the keV
neutrinos and eV ones since we have the ratio $m_{\nu keV}/m_{\nu eV}\approx10^{3}-10^{5}$.
Assuming that the seesaw mechanism is behind the neutrinos mass generation,
the eV scale for neutrino mass has its origin in the electroweak GeV
scale. In the same way, the keV scale may be related to new physics
at the TeV scale. The 3L3R model has such a new energy scale, $v_{\chi_{L}^{\prime}}$,
that makes it possible for the appearance of keV neutrinos mass
scale.

As for the right-handed neutrinos there are six of them acquiring
masses in the $\Lambda_{D}^{2}/\Lambda_{M}\approx10^{11}$~GeV scale
which, under our considerations, dictates the scale of the matrix
elements in $M_{R}$, and thus decoupled from the low energy spectrum.
These could be responsible for leptogenesis inducing the matter-antimatter
asymmetry in the observed Universe \cite{fukugita, sarkar}. Besides\textbf{
}explaining the observed neutrino oscillation with sub-eV masses,
the 3L3R model also gives us for free a potential candidate for the
dark matter of the Universe in the form of warm dark matter, namely,
the lightest $N_{L}$, and also heavy right-handed neutrinos suitable
to implement baryogenesis through leptogenesis.

Finally, it could be that with many competing scales present in the
model, some higher order effective operator violating baryon and lepton
numbers could engender proton decay in this model, as occurs in the
minimal $SU(3)_{L}\times U_{N}(1)$~\cite{truly}.\textbf{ }Such
effective operators involve four fermion multiplets with three of
them being quarks from the first family. The relevant lowest dimension
operator allowed by the gauge and discrete symmetries leads to the
four fermion operator \begin{align}
\frac{\xi}{\Lambda_{BL}^{3}}\varepsilon_{ijk}\varepsilon_{\alpha\beta\gamma}\overline{\left(Q_{1L}\right)_{i\alpha}^{c}}\left(Q_{1L}\right)_{j\beta}\overline{\left(\Psi_{1L}\right)_{\gamma}}\left(\eta_{R}^{T}\left(Q_{1R}\right)_{k}\right)\nonumber \\
\rightarrow\frac{\xi v_{\eta_{R}}}{\sqrt{2}\Lambda_{BL}^{3}}\varepsilon_{ijk}\varepsilon_{\alpha\beta\gamma}\overline{\left(Q_{1L}\right)_{i\alpha}^{c}}\left(Q_{1L}\right)_{j\beta}\overline{\left(\Psi_{1L}\right)_{\gamma}}\left(d_{1R}\right)_{k},\label{eq:oppd}\end{align}
where we define $i,j$, and $k$ as $SU(3)_{C}$ color indices, $\alpha,\beta$,and $\gamma$
as $SU(3)_{L,R}$ indices and $\xi$ a dimensionless constant of order
1. The suppression scale $\Lambda_{BL}$ is expected to be such
that \textbf{$\Lambda_{BL}>\Lambda_{D}\approx v_{\eta_{R}}$} once
it is related with baryon and lepton number violation. In accordance
with our previous assumption we can take $\Lambda_{BL}\equiv\Lambda_{M}$.
In this case the operator in Eq.$\left(\ref{eq:oppd}\right)$ is harmless
to the model. If, on the other hand, \textbf{$\Lambda_{BL}$} turns
out to be less than $\Lambda_{M}$ it is sufficient to have $\Lambda_{BL}>\left(M_{U}^{2}v_{\eta_{R}}\right)^{1/3}\approx10^{15}$
GeV, where $M_{U}\approx10^{16}$ GeV is a typical suppression factor
in grand unification theories. Although the model is not predictive
at this scale, concerning the proton decay no severe threat seems
to be posed by this process.

In summary, we developed a full LR symmetric gauge model for the electroweak
interactions based on the $SU(3)_{L}\times SU(3)_{R}\times U(1)_{X}$
symmetry.\textbf{ }The 3L3R model embeds $SU(3)_{L}\otimes U(1)_{N}$
model of Refs.~\cite{331nd,economical331model}, including the important
features well established in the Standard Model as well. Also, the model
offers an explanation for the family number puzzle, besides other
interesting phenomenological implications. It must be pointed out
that in this 3L3R model a peculiar neutrino scenario emerges interlacing
several open questions. There are three light neutrinos which may
explain the atmospheric and solar neutrino oscillation experiments; 
three with masses at keV range, which may provide a warm dark
matter candidate; and additional six heavy right-handed neutrinos
suitable to accommodate a baryogenesis through leptogenesis scheme.
This represents a logically consistent scenario not only for neutrino
physics, but also for the challenging problems of dark matter and
matter-antimatter asymmetry in the Universe, which may find in the
model presented here a self-contained explanation. These results give
to the 3L3R model enough strength to neatly represent a low energy
regime of a more fundamental underlying theory, showing that it is
of considerable interest to populate the desert between electroweak
and Planck scale.

This work was supported by CNPq (AGD, CASP and PSRS) and FAPESP (AGD).

\end{document}